\documentstyle[12pt]{article}
\begin{document}=
\def\Ket#1{|#1 \rangle}
\def\PP{\vskip.1in}
\def\mod{{\rm mod \,} }
\centerline{{\bf  Proof of the Impossibility of Non-Contextual}}
\centerline{{\bf  Hidden Variables in All Hilbert Space Dimensions}}
\vskip.15in
\centerline{Daniel I. Fivel}
\vskip.15in
\centerline{ Department of Physics }
\centerline{ University of Maryland, College Park, MD 20742-4111 }
\vskip.2in
\centerline{{\bf Abstract}}
\PP
It is shown that the algebraic structure of finite Heisenberg groups
associated with the tensor product of two Hilbert spaces leads to a
demonstration valid in all Hilbert space dimensions of the impossibility of
non-contextual hidden variables.
\PP
\PP
\PP
\PP
It has been known since the work of Bell and  Kochen and
Specker\cite{BELLa,KOCH}
that it is impossible to construct a non-contextual hidden-variable theory.
Such
theories would extend the quantum mechanical
characterization of a state in the following
way: A state would be characterized by a function $v$ assigning values to all
observables $a,b,\cdots$ (whether or not they commute) in such a way that (1)
$v$
assigns to each observable one of its eigenvalues, and (2) for any function
$f(c,d,\cdots)$ of a {\em commuting} subset of observables, the value assigned
must be
$f(v(c),v(d),\cdots)$. A discussion of the physical importance of ruling out
such theories is given in Mermin's review\cite{MERMa}.\PP

Recently there have been some enormous simplifications in the  proofs given by
Bell and
Kochen-Specker, the most elegant  being that of Peres\cite{PERa,PERb,MERMb}.
In common with the original proofs, however, the improved ones are specific to
particular
Hilbert space dimensions, and reveal no general underlying algebraic structure
responsible for the contradiction. The purpose of this paper is to provide a
proof that is
simultaneously valid in all Hilbert space dimensions and is directly linked to
the basic algebraic
structure of tensor products in Hilbert space.\PP

Let ${\cal H}$ denote an $N$ dimensional Hilbert space where $N$ is arbitrary
but will be fixed throughout the discussion.\PP

The Heisenberg group ${\cal G}_N$ may be defined abstractly as the group
generated by
$\sigma,\tau,I$ where $I$ is the identity and
$$
\sigma \tau = \omega\tau\sigma,\;\; \omega = e^{2\pi i/N}.
\eqno(1)
$$
It can be represented in ${\cal H}$ with basis
$\Ket{j}, \; j= 0,1,\cdots {\mod N}$ by the operators defined by:
$$
\sigma\Ket{j} = \omega^j\Ket{j},\; \tau\Ket{j} = \Ket{j+1},\;\;
{\rm with}\; \tau\Ket{N-1} = \Ket{0}.
\eqno(2)
$$
We write ${\cal H}_1$ and ${\cal H}_2$ to indicate two spaces of the same
dimension $N$. The spaces may be
thought of as being associated with two-particles of the same spin $J$ so that
$N = 2J+1$. Subscripts $1,2$ on operators indicate operators acting on the
respective particles. Now let $\sigma_1,\tau_1$ be a representation of the
Heisenberg group in a Hilbert space ${\cal H}_1$ and a similar pair for
particle-2. \PP

Next consider the following operators in the two-particle Hilbert space :
$$
A \equiv \sigma_1\otimes\tau_{2}^{-1},\;\; B = \tau_1\otimes\sigma_{2}^{-1},
\eqno(3)
$$
and the two operators in the spaces of particles 1 and 2:
$$
C = \sigma_1 \tau_1,\; D = \tau_2^{-1}\sigma_{2}^{-1}=(\sigma_2 \tau_2)^{-1}.
\eqno(4)
$$
{}From the Heisenberg commutation rule there follows:
$$
[A,B] = 0.
\eqno(5)
$$
Now observe that from the definitions we have the identity:
$$
 AB = C\otimes D,
\eqno(6)
$$
which has the structure of identities used by Peres\cite{PERa,PERb} and
Mermin\cite{MERMb}.\PP

We now attempt to implement a non-contextual assignment $v$ of values to
observables:\PP

Since the two factors in $A$ commute and the two factors of $B$ commute the
assignment hypothesis
applied to (6) leads to the identity:
$$
{{v(\sigma_1)v(\tau_1)}\over{v(\sigma_1\tau_1)}}=
{{v(\sigma_2)v(\tau_2)}\over{v(\sigma_2\tau_2)}}.
\eqno(7)
$$

But the two-particle state can be chosen arbitrarily, i.e.\ with particle-2
independent of
particle-1 . Hence the two sides must be independent of
the choice of state. Thus the ratio appearing in (7) must be independent of the
method of
assignment. In particular, since the eigenvalues of $\sigma,\tau,$ and
$\sigma\tau$ are the set of
$N$'th roots of unity, their complex conjugates are also eigenvalues. Hence if
$v$ is an assignment
with the required properties, so also is its complex conjugate. Hence the
common value in $(7)$ must
be {\em real}. \PP

The derivation of (7) made use only of the Heisenberg commutation rule. This
means that
 choices of the $\sigma$ and $\tau$ for the two spaces can be made
independently. Now
suppose ${\cal U}$ is an arbitrary {\em anti-unitary} transformation on ${\cal
H}$ and let us define
$$
\sigma^{\prime} = {\cal U}\tau{\cal U}^{-1}, \;\; \tau^{\prime} = {\cal
U}\sigma{\cal U}^{-1}.
\eqno(8)
$$
Because $\omega$ is unimodular the anti-unitarity of ${\cal U}$ insures that
the primed operators
have the same Heisenberg commutation relation as the unprimed operators. If we
choose a particular
representation $\sigma,\tau$ for particle-1, we shall now fix the the
representation for particle-2
to be $\sigma^{\prime},\tau^{\prime}$ as defined by $(8)$ for some arbitrary
but fixed anti-unitary
operator ${\cal U}$.\PP

Our next step will be to use the selected ${\cal U}$ to construct a certain
distinguished
two-particle state associated with it:\PP

 For any state $\Ket{x}$ introduce the shorthand:
$$
\Ket{x^{{\cal U}}} \equiv {\cal U}\Ket{x},
\eqno(9)
$$
so that if $\Ket{x}$ is an eigenstate of an operator $\Lambda$ with eigenvalue
$\lambda$,
then  $\Ket{x^{\cal U}}$ is an eigenstate of ${\cal U}\Lambda{\cal U}^{-1}$
with
eigenvalue $\lambda^*$. Now define the required two-particle state by:\PP

$$\Ket{{\cal U}} = N^{-1/2}\sum_{n=1}^{N}{\Ket{n,1}\otimes\Ket{n^{{\cal
U}},2}}.
 \eqno(10) $$
Note that our use of the same letter ${\cal U}$ to label both the state and
the operator that appears on the right is justified by the fact that
 the right side is independent
of the choice of basis. To see this observe that if $\Ket{\tilde{n}}$ is
another basis related to
the $\Ket{n}$ basis by a unitary matrix $\alpha$, i.e.\
$$
\Ket{n} = \sum_j{  \alpha_{nj} \Ket{ \tilde{j} }},
\eqno(11)
$$
then the antiunitarity of ${\cal U}$ and the unitarity of $\alpha$ lead to the
conclusion that (10)
is unchanged under the replacement of $\Ket{n,\nu}$ by $\Ket{ \tilde {n},\nu},
\; \nu = 1,2.$ This
basis independence makes (10) a  so-called ``perfectly entangled"
two particle state, i.e.\ it has the property that if particle-1 is found in
any state $\Ket{x}$
then its partner will be found with certainty in the state $\Ket{x^{{\cal
U}}}$. In particular if
particle-1 is found in an eigenstate of $\sigma $ with eigenvalue $\lambda$,
then particle-2 will be
found with certainty in an eigenstate of ${\cal U}\sigma{\cal U}^{-1} =
\tau^{\prime}$ with
eigenvalue $\lambda^*$. A similar remark holds for $\tau$. If particle-1 is
found
in an eigenstate of $\sigma \tau$ with eigenvalue $\lambda$, then its partner
will be found in an
eigenstate of
$$
{\cal U}\sigma\tau{\cal U}^{-1} = \tau^{\prime}\sigma^{\prime} =
\omega^{-1}\sigma^{\prime}\tau^{\prime},
\eqno(12)
$$
with eigenvalue $\lambda^*$.\PP
It thus follows that for a two particle system in the state $\Ket{{\cal U}}$ a
non-contextual
assignment $v$ must satisfy:
 $$
v(\sigma_2) = v^*(\tau_1),\; v(\tau_2) = v^*(\sigma _1).
$$
and
$$
\omega^{-1}v(\sigma_2\tau_2) = v^*(\sigma_1\tau_1).
\eqno(13)
$$
Thus there follows from (7) and the reality of the two sides:
$$
\omega = 1,
\eqno(14)
$$
which is false for all $N$. Thus we have established a contradiction and proved
the impossibility
of a non-contextual theory for all $N$.\PP

The advantage of this proof is that it exposes some of the source of the
algebraic contradictions.
While it does not reveal the significance of the 117 directions in
Kochen-Specker's original
demonstration, it does reveal why the case $N=2$ allows a special
kind of demonstration. Indeed the following connection can be made to Peres'
and Mermin's
examples which use the spin operators: Because $\omega$ is real for $N=2$ (but
no other $N$) there
is a special anti-unitary transformation of the Heisenberg generators, namely
$\sigma \to -\sigma$ and $\tau \to -\tau$. Under this transformation we see
that $\mu \equiv
-i\sigma\tau$ transforms according to $\mu \to -\mu$. But the three
operators $\sigma,\tau,\mu$ now have the commutator algebra of the Pauli
matrices {\it in
consequence of the Heisenberg algebra} (for $N=2$ only) so that the algebraic
contradictions
resulting from the spin algebra can also be regarded as a manifestation  of the
Heisenberg
structure.\PP

While the proof relied on the properties of an entangled state, the choice of a
particular
entangled state was completely arbitrary. Moreover, one knows \cite{FIV} that
these states
{\em span} the two-particle Hilbert space. Furthermore the construction of
$\Ket{{\cal U}}$ was
canonically related to the isomorphism that interchanges $\sigma$ and $\tau$.
Since Heisenberg
groups can actually be  {\em defined} by the property of isomorphism between a
translation group (in
this case on the lattice ${\rm mod} \; N$) and its Pontryagin dual (i.e.\ its
character group), the
introduction of $\Ket{{\cal U}}$ cannot be considered a ``deus ex machina".
Rather one may look at
the above proof as revealing a new
 way in which entanglement characterizes the most dramatically non-classical
properties of quantum
mechanics. Such  properties are of considerable current interest in connection
with their
potential application to the securing of communication
channels\cite{FIV,BEN,WOOT}   Moreover, since
entangled states also enter into the other Bell no-hidden-variable
theorem\cite{BELLb} (predicated on
locality), one anticipates a linking of non-contextuality and non-locality at a
fundamental level.
In particular the fact\cite{MERMb,GREEN} that there are examples that serve
simultaneously as
counter-examples to both types of no-hidden-variable theorem is not too
surprising. \PP
\PP

\end{document}